\documentclass[%
 reprint,
 amsmath,amssymb,
 aps,prl,onecolumn
]{revtex4-2}

\usepackage{graphicx}%
\usepackage{dcolumn}%
\usepackage{multirow}
\usepackage{amssymb}
\usepackage{comment}
\usepackage{MnSymbol,wasysym}
 \usepackage{setspace}

\usepackage[dvipsnames,table,xcdraw]{xcolor}

\usepackage{hyperref}
\hypersetup{
    colorlinks=true,
    linkcolor=Blue,
    filecolor=Blue,      
    urlcolor=Blue,
    citecolor=blue,
}

\definecolor{Gray}{gray}{0.85}
\newcolumntype{g}{>{\columncolor{Gray}}r}
\newcolumntype{w}{>{\columncolor{white}}r}

\renewcommand\vec[1]{\ensuremath\boldsymbol{#1}} 

\begin{document}

\title{
Superconductivity-induced improper orders
}

\author{Andr\'{a}s L. Szab\'{o}}
\affiliation{Institute for Theoretical Physics, ETH Zurich, 8093 Zurich, Switzerland}

\author{Aline Ramires}
\affiliation{Paul Scherrer Institute, 5232 Villigen PSI, Switzerland}

\date{\today}

\begin{abstract}
The study of improper phases in the context of multiferroic materials has a long history, but superconductivity has yet to be connected to the network of ferroic orders. 
In this work, we highlight an overlooked mechanism that couples superconducting order parameters to odd-parity orders in
the charge or spin sectors such that the latter emerge as improper orders. 
For that, we explore a novel perspective of nonsymmorphic symmetries based on extended symmetry groups in real space.
We highlight how nonsymmorphic symmetries can generate rather nonintuitive couplings between order parameters. 
In particular, we find that a bilinear in the superconducting order parameter can couple linearly to odd-parity orders in centrosymmetric systems. 
Our findings can account for the unusual phenomenology of CeRh$_2$As$_2$, a recently discovered heavy fermion superconductor, and open the door for exploring nonsymmorphic symmetries in the broader context of improper orders with potential applications to functional materials.
\end{abstract}

\maketitle 


The Landau theory of phase transitions has been a leading framework for understanding ordered phases of matter.
It has been successfully applied to describe magnetic and electric ordering and their non-trivial interplay in multiferroic systems, which are central in the pursuit of exotic functional materials \cite{Fiebig2016, Nordlander2019}.
Within multiferroics, improper ferroelectrics 
develop electric polarization controlled by the development of a leading distortive or magnetic order parameter \cite{Levanyuk1974, Dvok1974, Tokura2014, Benedek2022}. 
More generally, improper phases are associated with order parameters that develop as a secondary effect of the development of a leading order. 
The interplay of superconductivity with magnetic and charge orders is empirically known in multiple families of materials and extensively discussed in the framework of intertwined \cite{Fradkin2015} or vestigial orders \cite{Fernandes2019}. Nevertheless, their relation in the context of improper orders has not yet been fully investigated, as the complexity of superconducting order parameters has only recently started to be acknowledged. 
In this work, we explore the untapped realm of superconducting-induced improper orders, highlighting the role of nonsymmorphic symmetry for the development of unexpected couplings between order parameters, which can potentially explain the unusual phenomenology recently reported in a material in the family of heavy fermion systems.







\begin{figure}[t]
    \centering
    \includegraphics[width=9cm]{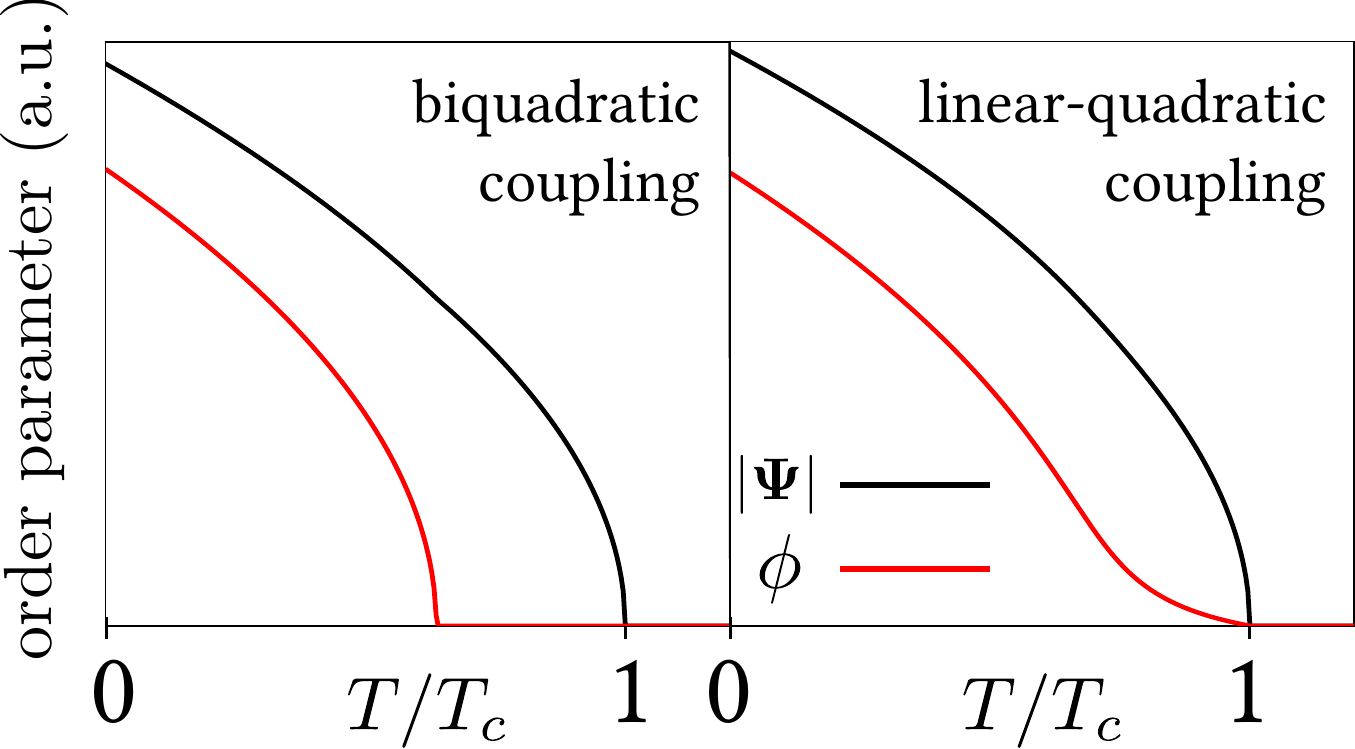}
    \caption{
    Order parameter coupling within Landau theory. Assuming a primary superconducting order parameter $\mathbf{\Psi}$ (potentially multicomponent), and a secondary order $\phi$, this figure highlights the distinct phenomenology that emerges from two different types of coupling between them.
    Left: For a biquadratic coupling, $f_\lambda = \lambda |\phi|^2|\mathbf{\Psi}|^2$, there are two transition temperatures and both order parameters onset with the characteristic $\sqrt{T}$ temperature dependence. 
    Right: for linear-quadratic coupling, $f_\lambda \sim \lambda \phi (\Psi_1^*\Psi_2 \pm \Psi_1\Psi_2^*)$, there is a single transition temperature and the subleading order parameter $\phi$ onsets with a slower, linear temperaure dependence. Here $\lambda$ is a generic coupling constant. More details are given in the SI. 
    }
    \label{fig:GL}
\end{figure}

The development of ordered phases of matter can be 
understood on the phenomenological level based on the notion of symmetry breaking associated with the onset of an 
 order parameter.
In crystalline solids, the primary symmetries involved are spatial, generally accompanied by time-reversal symmetry. 
Spatial symmetries include translations, rotations, and reflections, as well as combinations of these.
Particularly notable are nonsymmorphic systems,  as these feature symmetry transformations that are necessarily accompanied by a fractional primitive lattice vector (PLV) translation.
These systems have been extensively explored in the context of topological band structures \cite{Mong2010,Liu2014,Zhao2016, Chen2016,Bradlyn2016}, and such symmetries are key to protecting band degeneracies of Bloch's states with opposite parity at specific points in momentum space~\cite{Dresselhaus2008, Bradley2010}. 
However, the effects of nonsymmorphic symmetries in the context of ordered states of matter are less explored, and most efforts have been focused on their topological classification \cite{Wang2016, Shiozaki2016,Kobayashi2016}. 
Similarly, much of the research in the context of superconductivity relies on the analysis of point group symmetries, the crystalline symmetries that are left once one factors out translations, a trivial procedure in symmorphic systems.
Nonsymmorphic symmetries complicate the process of factoring out translations, making the group theoretical analysis and the classification of superconducting states more cumbersome, particularly if one works in a momentum space representation \cite{Bradley2010, Cvetkovic2013}. 
Here, we take a complementary view of nonsymmorphic symmetries in the context of ordered phases of matter. 
We classify the order parameters in the superconducting, charge, and spin sectors based on their textures directly in real space \cite{Vanderbos2016}, taking explicit account of the nonsymmorphic nature of the crystal \cite{PerezMato2016}. 
From this analysis, we find new types of coupling between superconducting order parameters and orders in the spin or charge sectors, which can induce odd-parity improper orders by the development of a primary order in the superconducting channel, see Fig.~\ref{fig:GL}.
Our results could lead to new functionalities and technological applications by highlighting an unapped  mechanism for the development of improper orders and bringing new connectivities between superconductivity and other functional phases of matter.

\begin{figure*}[t]
    \centering
    \includegraphics[width=12cm]{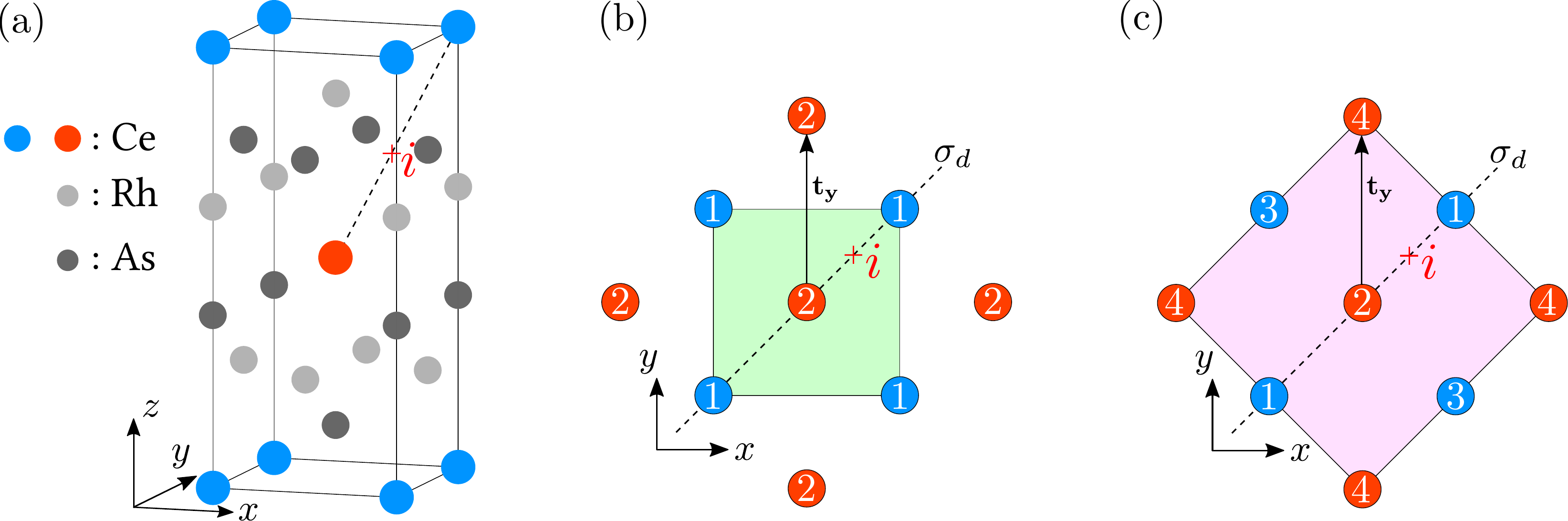}
    \caption{(a) Crystal structure of CeRh$_2$As$_2$, with centrosymmetric nonsymmorphic space group $P4/nmm$ ($\#$ 129). The Ce atoms span a body-centered tetragonal lattice with the main fourfold rotation axis going through these sites. The crystal field due to the Rh and As atoms (dark and light grey spheres, respectively) breaks inversion symmetry at the Ce sites, generating a Ce sublattice structure (blue and red spheres), characterizing this system as locally noncentrosymmetric. Notably, the inversion center is located at the midpoint between Ce sublattices, 
    (b) Original unit cell (green) projected to the $x-y$ plane containing two Ce atoms (1 and 2) and (c) enlarged unit cell (magenta) containing four Ce atoms (1 through 4). Dashed line represents the diagonal mirror plane, red cross is the global inversion centre, coinciding with our chosen origin. Translation by one lattice constant (indicated by $\mathbf{t_y}$) is a trivial operation in the original unit cell scenario, as it takes a sublattice into itself. In contrast, in the enlarged unit cell scenario a translation by one lattice constant constitutes a new operation.}
    \label{fig:unit_cell}
\end{figure*}

\begin{figure*}[t]
    \centering
    \includegraphics[width=18cm]{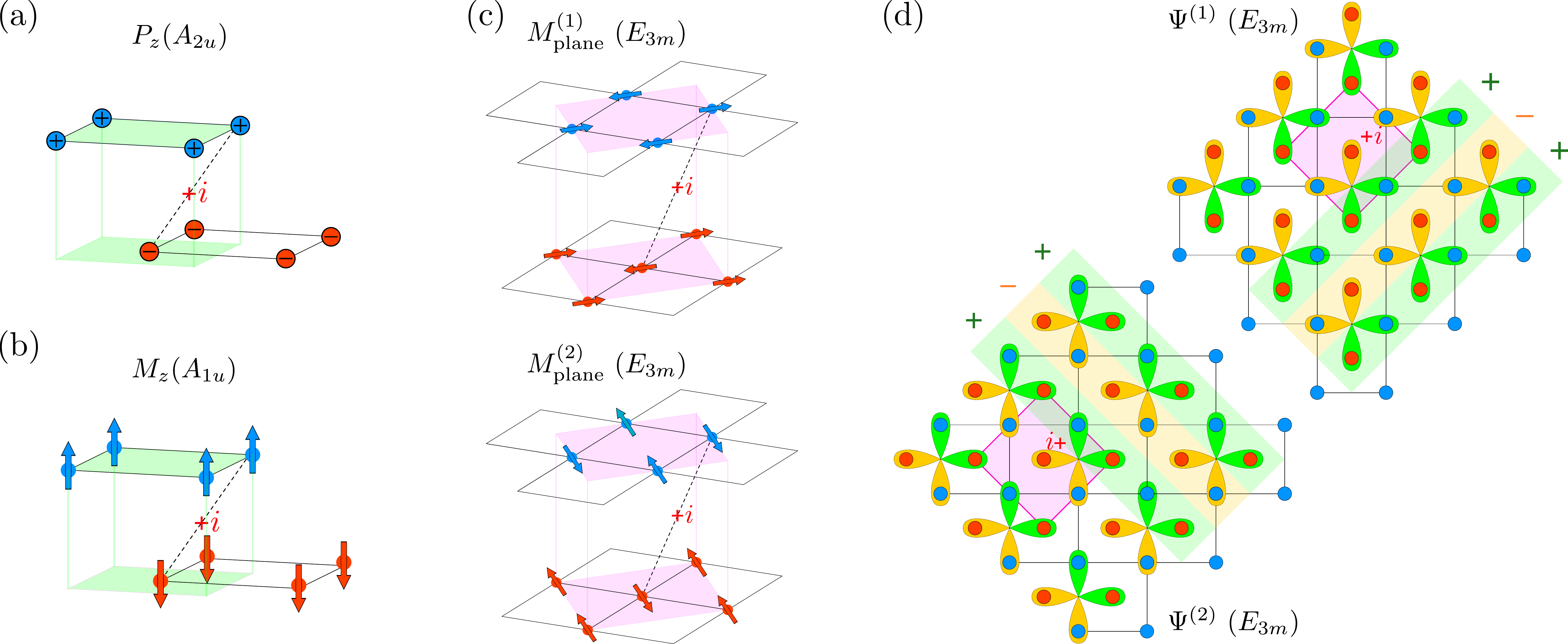}
    \caption{Representative orders. Inversion-odd $P_z$ polar (a) and $M_z$ layer antiferromagnet (b) orders, adhering to the original unit cell (green). (c) Two components of $E_{3m}$ in-plane magnetic order (d) two components of $E_{3m}$ superconducting order in the enlarged unit cell (magenta). The pairing wave function is intrasublattice but antisymmetric between sites, resulting in a modulation in the projected x-y plane ($+/-$, marked by green and orange respectively). In (c) and (d) the components are related to each other via 90 degree rotation, but one component is odd-parity, while the other is even. }
    \label{fig:orders}
\end{figure*}

The recently discovered heavy-fermion compound CeRh$_2$As$_2$, depicted in Fig. \ref{fig:unit_cell} (a), realizes an exotic phase diagram with two superconducting phases as a function of a $c$-axis magnetic field \cite{Khim2021}. 
While the pairing symmetry of the two phases is to date unclear, nuclear magnetic resonance (NMR)~\cite{Ogata2023} and nuclear quadrupole resonance (NQR)~\cite{Kibune2022} measurements on the As sites reveal antiferromagnetic order coexisting with the low-, but not with the high-field superconducting phase.
Most intriguingly, specific heat and thermal expansion measurements indicate a single phase transition~\cite{Semeniuk2023}, suggesting that the onset of superconductivity coincides with that of magnetism in the entire low-field phase, in a range of magnetic fields spanning 4~T.
Both NMR and NQR measurements observed site-selective broadening of the signal within the magnetic phase: of the two inequivalent As sites in the unit cell, only one of them experiences a change in the local magnetic field with the onset of magnetic ordering.
These observations constrain the possible textures for magnetic moments localized at the Ce sites.
Within homogeneous phases, the only consistent order is a layered antiferromagnet, with magnetic moments along the $c$-axis, changing sign across the sublattices [see Figs. \ref{fig:unit_cell} (b) and \ref{fig:orders} (b)].
Two further families of solutions with in-plane magnetic moments can be obtained by doubling the unit cell [see Figs. \ref{fig:unit_cell} (c) and \ref{fig:orders} (c)]. 
Below, we discuss how these constraints on the magnetic order impose strong restrictions on the nature of the superconducting state in this system.



Within the Landau formalism, the coexistence of superconducting and magnetic orders implies a coupling between the corresponding order parameters consistent with all operative symmetries.
The most straightforward gauge-invariant coupling  preserving time-reversal symmetry is quadratic in both the superconducting ($\Psi$) and magnetic ($M$) order parameters, e.g. $\sim M^2 |\Psi|^2$.
This coupling, however, does not result in the same critical temperatures for the two phases.
From the phenomenology of improper orders, a linear-quadratic coupling between $M$ and $\Psi$ with a dominant superconducting order would lead to the onset of magnetism at the superconducting critical temperature (see Fig. \ref{fig:GL} and discussion in the SI). For this type of coupling, we need to recourse to a multi-component superconducting order parameter, $\boldsymbol{\Psi} = (\Psi_1,\Psi_2)$, which would allow for a term $\sim iM(\Psi_1 \Psi_2^\ast-\Psi_1^\ast \Psi_2)$ in the free energy.

In CeRh$_2$As$_2$, the homogeneous magnetic order is odd parity, as the magnetic moments are opposite for sites related by inversion symmetry [see Fig. \ref{fig:orders} (b) and (c)]. Due to the global inversion symmetry, order parameters classified from the perspective of point group symmetries are generally labeled as either even or odd, and superconducting order parameter bilinears are invariably even parity.
This simplified view naively makes a linear-quadratic coupling between magnetic and superconducting order parameters impossible in CeRh$_2$As$_2$, as it would require 
$\Psi_1$ and $\Psi_2$ to be of opposite parity.
Below, we show that in the absence of accidental degeneracies, modulated superconducting orders in nonsymmorphic systems can sustain multi-component order parameters with opposite parity, allowing for this unusual coupling and for the emergence of magnetism as an improper order triggered by a primary superconducting order.







More concretely, taking as a working example the space group $P4/nmm$ ($\#129$), we now discuss how unusual irreducible representations (irreps) with components of opposite parity emerge once we enlarge the symmetry group by extending the unit cell to account for modulated order parameters.
The 16 symmetry operations in $P4/nmm$ are generated by $\bar{C}_{4z}=\{C_{4z}|\mathbf{t_{x}}/2\}$, a rotation by $\pi/2$ along the z-axis followed by $\mathbf{t_{x}}/2$, half a PLV translation along the x-axis; $\sigma_d=\{\sigma_d|\mathbf{0}\}$, a mirror reflection along the diagonal plane ($x=y$); and inversion $ i = \{i|\mathbf{0}\}$ (the complete list of group operations in the standard Seitz notation is given in the SI).
Given the nonsymmorphic nature of the space group, these 16 symmetry operations, when composed, do not close into themselves. 
If we are interested in homogeneous phases, we can redefine the composition of these operations modulo  integer lattice vector translations such that they form a group. 
This procedure corresponds to factoring out translations to determine the little group at the $\Gamma$ point in momentum space \cite{Bradley2010}, which is isomorphic to $D_{4h}$. For this group, the symmetry elements are organized in 10 conjugacy classes, leading to 10 irreps which are labeled as even or odd parity (see details in the SI).

If we allow the system to develop modulations encompassing multiple unit cells in a 
commensurate manner, we need to consider the corresponding wave-vector dictating the modulation. 
If the wave vector corresponds to a point in momentum space at the edge of the BZ, we expect unusual degeneracies in nonsymmorphic systems, as is extensively discussed in the context of electronic band structures in momentum space. 
Choosing the simplest scenario, here we double the unit cell and introduce the notion of an \emph{extended group} by adding to the original group new symmetry operations, which take one primitive unit cell into another in the doubled unit cell \cite{Vanderbos2016}; see Fig. \ref{fig:unit_cell} (b) and (c). 
The extended symmetry group is formed by the original 16  operations plus the composition of these with a PLV translation, here chosen to be $E' = \{E|\mathbf{t_y}\}$, which is the extension of the identity operation $E=\{E|\mathbf{0}\}$ (extended symmetry operations are denoted with a prime). Extending the group of symmetries to 32 elements leads to novel irreps (see SI for details). 
If the space group is symmorphic, the total number of irreps is simply doubled, and the new irreps behave as the original ones up to an extra minus sign under operations including a 
PLV translation.
In contrast, if the group is nonsymmorphic, the irreps associated with the extended group can be fundamentally distinct from the original irreps.

To understand how nontrivial irreps are generated, we rely on two elementary results from group theory: (i) the number of irreps is equal to the number of conjugacy classes; (ii) the dimensions of irreps should follow the identity $\sum_i |d_i|^2 = |\mathbf{G}|$, where $d_i$ is the dimension of the $i$-th irreducible representation  and $|\mathbf{G}|$ is the order of the group (the number of elements in the group). 
The order of the extended group in our example is twice the order of the original group. For symmorphic systems, the number of irreps of a given dimension is also doubled, trivially satisfying points (i) and (ii). 
For nonsymmorphic systems, the conjugacy classes in the extended group are  not twice as many as in the original group. 
This happens because some of the original and extended operators necessarily coalesce into a single conjugacy class. 
This point can be understood by considering the conjugation of a generic spatial symmetry operation $O_B = \{R_B| \mathbf{t}_B\}$ by another generic operation $O_A = \{R_A| \mathbf{t}_A\}$:
\begin{eqnarray}
    \label{Eq:Conjugation}
O_A^{-1}.O_B.O_A
    = O_C= \{R_A^{-1}R_BR_A|R_A^{-1}(R_B\mathbf{t}_A-\mathbf{t}_A+\mathbf{t}_B)\},
\end{eqnarray}
where the dot denotes the composition of operations,  
 and we used $O_A^{-1} = \{R_A^{-1}| -R_A^{-1}\mathbf{t}_A\}$. The presence of inversion operation in P4/nmm allows us to choose $O_A = \{i|\mathbf{0}\}$, such that the RHS of the equation above reads $\{R_B|-\mathbf{t}_B\}$.
 In the original symmetry group, associated with the primitive unit cell, if $O_B$ is nonsymmorphic with $\mathbf{t}_B$ corresponding to half a PLV translation, we can redefine $-\mathbf{t}_B = \mathbf{t}_B+PLV \equiv \mathbf{t}_B$. As a consequence, $O_C=O_B$ and we do not get any information about conjugacy classes. 
 On the other hand, in the extended symmetry group $-\mathbf{t}_B = \mathbf{t}_B+PLV \not\equiv \mathbf{t}_B$, such that $O_C = O_B' \neq O_B$. 
 The relation $O_A^{-1}.O_B.O_A = O_B'$ indicates that $O_B$ and $O_B'$ belong to the same conjugacy class in the extended group. 
 We conclude that all conjugacy classes containing nonsymmorphic operations in the original symmetry group are enlarged in the extended symmetry group.
 This coalescence of conjugacy classes tells us that we have less than twice as many irreps in the extended group, and, consequently,  the new (double-valued) irreps are generally not one-dimensional. 
 The complete analysis is summarized in Table \ref{Tab:Character} and a more detailed discussion is provided in the SI.
 In this example, there are 14 conjugacy classes, therefore 14 irreps. 
 There are 4 new two-dimensional irreps in the extended group (labelled as $E_{im}$, $i=1,...,4$, with $m$ standing for mixed parity), with the unusual property of having zero character associated with inversion symmetry.

\begin{table}
  \begin{center}
\begin{tabular}{c|
>{\centering\columncolor[gray]{0.9}}m{0.8cm}
>{\centering\columncolor[gray]{0.9}}m{0.8cm}
>{\centering\columncolor[gray]{0.9}}m{0.8cm}
>{\centering\columncolor[gray]{0.9}}m{0.8cm}
>{\centering} m{0.8cm}
>{\centering} m{0.8cm}
>{\centering\columncolor[gray]{0.9}}m{0.8cm}
>{\centering\columncolor[gray]{0.9}}m{0.8cm}
>{\centering} m{0.8cm}
>{\centering} m{0.8cm}
>{\centering} m{0.8cm}
>{\centering} m{0.8cm}
>{\centering\columncolor[gray]{0.9}}m{0.8cm}
>{\centering\arraybackslash\columncolor[gray]{0.9}}m{0.8cm}
}
$G_{16}^9$ &\multicolumn{2}{c}{$E$}& \multicolumn{2}{c}{$\tilde{C}_{2z}$} &$ 2\bar{C}_{4z}$ & $2 \bar{\sigma}_x $&  \multicolumn{2}{c}{$2{\sigma}_d$} &$ i $&$ \tilde{\sigma}_h$ & $2 \bar{S}_4 $& $2 \bar{C}_{2x}$ & \multicolumn{2}{c}{$2 {C}_{2\bar{d}}$}
 \\ 
 &\multicolumn{2}{c}{$\swarrow\searrow$}& \multicolumn{2}{c}{$\swarrow\searrow$} & $\Downarrow$ & $\Downarrow$ &  \multicolumn{2}{c}{$\swarrow\searrow$} & $\Downarrow $& $\Downarrow$ & $\Downarrow$ & $ \Downarrow $ & \multicolumn{2}{c}{$\swarrow\searrow$}
 \\ 
\rowcolor{white} $G_{32}^2$  & $E$ & 
$E'$ & 
$\tilde{C}_{2z} $&
$\tilde{C}_{2z}'$ &
$4\bar{C}_{4z}$ &
$4\bar{\sigma}_x$ &
$2{\sigma}_d$ &
$2{\sigma}_d'$ &
$2i$ & 
$2\sigma_h$ &
$4 \bar{S}_4$ & 
$4 \bar{C}_{2x}$ & 
$2 {C}_{2\bar{d}}$& 
$2 {C}_{2\bar{d}}'$ \\ \hline
$A_{1g}$  & 1 & 1  & 1  & 1  & 1   & 1  & 1  & 1  & 1  & 1  & 1 & 1 & 1 & 1\\
$A_{2g}$ & 1 & 1  & 1  & 1  &  1  & \hspace{-0.1cm}-1 & \hspace{-0.1cm}-1 &\hspace{-0.1cm}-1 & 1  & 1 & 1 &\hspace{-0.1cm}-1 & \hspace{-0.1cm}-1 &\hspace{-0.1cm}-1\\
$B_{1g}$ & 1 & 1  & 1  & 1  & \hspace{-0.1cm}-1  &  1 & \hspace{-0.1cm}-1 & \hspace{-0.1cm}-1 & 1  & 1  & \hspace{-0.1cm}-1 & 1 &\hspace{-0.1cm}-1 &\hspace{-0.1cm}-1\\
$B_{2g}$ & 1 & 1  & 1  & 1  &\hspace{-0.1cm}-1  &\hspace{-0.1cm}-1 & 1  & 1  & 1  & 1  & \hspace{-0.1cm}-1 & \hspace{-0.1cm}-1 & 1 & 1\\
$E_{g} $& 2 & 2  & \hspace{-0.1cm}-2 &\hspace{-0.1cm}-2 & 0   &  0 & 0  & 0  & 2  & \hspace{-0.1cm}-2 & 0 & 0 & 0 & 0\\ 
$A_{1u} $& 1 & 1  & 1  & 1  & 1   & \hspace{-0.1cm}-1 & \hspace{-0.1cm}-1 & \hspace{-0.1cm}-1 & \hspace{-0.1cm}-1 & \hspace{-0.1cm}-1 & \hspace{-0.1cm}-1 & 1 & 1 & 1\\
$A_{2u} $& 1 & 1  & 1  & 1  &  1  & 1  & 1  & 1  & \hspace{-0.1cm}-1 &\hspace{-0.1cm}-1 &\hspace{-0.1cm} -1 & \hspace{-0.1cm}-1 & \hspace{-0.1cm}-1 & \hspace{-0.1cm}-1\\
$B_{1u} $& 1 & 1  & 1  & 1  & \hspace{-0.1cm}-1  & \hspace{-0.1cm}-1 & 1  & 1  & \hspace{-0.1cm}-1 & \hspace{-0.1cm}-1 & 1 & 1 & \hspace{-0.1cm}-1 &\hspace{-0.1cm}-1\\
$B_{2u} $& 1 & 1  & 1  & 1  & \hspace{-0.1cm}-1  & 1  &\hspace{-0.1cm}-1 & \hspace{-0.1cm}-1 & \hspace{-0.1cm}-1 & \hspace{-0.1cm}-1 & 1 &\hspace{-0.1cm}-1 & 1 &1\\
$E_{u} $& 2 & 2  & \hspace{-0.1cm}-2 & \hspace{-0.1cm}-2 & 0   & 0  & 0  & 0  & \hspace{-0.1cm}-2 & 2  & 0 & 0 & 0 & 0\\ \hline
\rowcolor{white}$E_{1m} $& 2 & \hspace{-0.1cm}-2 & 2  &\hspace{-0.1cm}-2 & \cellcolor{Goldenrod} 0   & \cellcolor{Goldenrod} 0  & 2  & \hspace{-0.1cm}-2 & \cellcolor{Goldenrod}0  & \cellcolor{Goldenrod}0 & \cellcolor{Goldenrod}0 & \cellcolor{Goldenrod}0 & 0 & 0\\
\rowcolor{white}$E_{2m} $& 2 & \hspace{-0.1cm}-2 & 2  & \hspace{-0.1cm}-2 & \cellcolor{Goldenrod}0   &\cellcolor{Goldenrod} 0  & \hspace{-0.1cm}-2  & 2  & \cellcolor{Goldenrod}0  & \cellcolor{Goldenrod}0 & \cellcolor{Goldenrod}0 & \cellcolor{Goldenrod}0 & 0 & 0\\
\rowcolor{white}$E_{3m} $& 2 &\hspace{-0.1cm}-2 & \hspace{-0.1cm}-2 & 2  & \cellcolor{Goldenrod}0   & \cellcolor{Goldenrod}0  & 0  & 0  & \cellcolor{Goldenrod}0  &\cellcolor{Goldenrod}0 & \cellcolor{Goldenrod}0 &\cellcolor{Goldenrod} 0 & 2 & \hspace{-0.1cm}-2\\
\rowcolor{white}$E_{4m} $& 2 & \hspace{-0.1cm}-2 & \hspace{-0.1cm}-2 & 2 & \cellcolor{Goldenrod}0    &\cellcolor{Goldenrod} 0  & 0  & 0  &\cellcolor{Goldenrod} 0  &\cellcolor{Goldenrod} 0 &\cellcolor{Goldenrod} 0 &\cellcolor{Goldenrod} 0 &\hspace{-0.1cm}-2 & 2
\end{tabular}
\end{center}
\caption{Character table for P4/nmm modulo two integer lattice translations. 
The first line gives the 10 conjugacy classes for P4/nmm modulo integer lattice translations (isomorphic to $D_{4h}$ and the abstract group $G_{16}^9$ \cite{Bradley2010}). 
The second line encodes the 14 conjugacy classes in P4/nmm modulo two integer lattice translations (isomorphic to abstract group $G_{32}^2$\cite{Bradley2010}). 
The symmetry operations are labeled according to their associated point group operation. If the operation is accompanied by a half PLV translation, it is marked with a bar, while if it is accompanied by two orthogonal half PLV translations, it is marked with a tilde. Operations  without either a bar or a tilde are pure point operations. Operations marked with a prime belong to the set of operations extended by a PLV translation.
The arrows indicate if the original conjugacy class splits ($\swarrow\searrow$) or doubles ($\Downarrow$) in the extended group. 
The first 10 irreps have well-defined parity, and their labels follow that of the $D_{4h}$ point group. 
The 4 last irreps are new to the extended group and have mixed parity. 
The subscripts correspond to even (g), odd (u), or mixed (m) parity. In gray color, we highlight the columns that are simply doubled for the initial 10 irreps given the splitting on the conjugacy classes. In yellow, we highlight the columns with zero characters due to the coalescence of the original and extended operations in the same conjugacy class.
}
\label{Tab:Character}
\end{table}

To better understand this last statement, we can conjugate inversion with a nonsymmorphic operation. Choosing $O_B = \{i|\mathbf{0}\}$, we find the RHS of Eq.~\ref{Eq:Conjugation} reads $\{i|-2R_A^{-1}\mathbf{t}_A\}$.
If $O_A$ is nonsymmorphic and $\mathbf{t}_A$ is half a PLV, $-2R_A^{-1}\mathbf{t}_A$ is a PLV, and therefore nonsymmorphic group elements inevitably connect inversion $i = \{i|\vec{0}\}$ with $i' = \{i|\mathbf{t_y}\}$, leading to the enlargement of the associated conjugacy class under the group extension.
This result has strong implications. The fact that the counjugacy class associated with the inversion operation is enlarged tells us that the character associated with the new irreps is zero \cite{Bradley2010}. As a consequence, the new two-dimensional irreps are associated with two-component basis functions with components of opposite parity. 
This fact is directly associated with known results in electronic band theory: our extended group is isomorphic to the abstract group $G_{32}^2$, the little group at the $M$ point \cite{Bradley2010}.
This remarkable result challenges our intuition about inversion symmetry. In centrosymmetric systems with nonsymmorphic symmetries inversion symmetry can be ``effectively broken"  if the system develops textures with wave-vectors that lie at the edge of the BZ.


 
Going back to the discussion on CeRh$_2$As$_2$, using the original group structure modulo PLV (the little-group at the $\Gamma$ point), we can classify the order parameters in the charge, spin, and superconducting sectors in terms of irreps, which are either even or odd under parity (details in SI). 
In contrast, if we classify the order parameters according to extended group (see details in SI), we find orders associated with irreps labeled as $E_{im}$ 
having two components that transform differently under inversion.
Curiously, the magnetic order with in-plane moments discussed above belongs to the irrep $E_{3m}$, but it cannot couple to any superconducting order through the desired linear-quadratic term (see discussion in SI). 
On the other hand, the magnetic order with moments along the $c$-axis is associated to the irrep $A_{1u}$, which is odd parity, and can couple linearly to quadratic terms in the superconducting order parameter if the latter is associated with irreps $E_{3m}$ or $E_{4m}$. 
The latter type of coupling has strong implications for the phenomenology. Magnetism develops as an improper order, and its temperature dependence does not follow the standard $\propto \sqrt{T}$ behaviour expected for a leading order parameter within Landau theory, as depicted in Fig. \ref{fig:GL}. 
Improper orders onset with a weaker temperature dependence $\propto T$, which might make their experimental observation more difficult. 
A slow onset for the improper magnetic order in CeRh$_2$As$_2$ is consistent with the apparent magnetic critical temperature being lower than the superconducting critical temperature \cite{Kibune2022,Ogata2023}. These results should trigger a more in-depth investigation to determine the temperature at which magnetism emerges. If magnetism onsets exactly at $T_c$, our analysis suggests that we have strong constraints to the nature of both magnetism and superconductivity. It necessarily indicates that the magnetic moments are aligned along the z-axis and that superconductivity is of a very unusual type: a chiral PDW 
with two components of opposite parity.



Classifying order parameters in real space taking into account nonsymmorphicity by dealing with extended symmetry groups allows us to systematically account for modulated unconventional orders and novel types of couplings between them. 
Nonsymmorphic symmetries are related to intrinsically complex crystalline structures.
Nonsymmorphic crystals necessarily have sublattice structures that add to the already complex set of charge, orbital, and spin degrees of freedom necessary to faithfully describe the electronic structure of most materials. 
The richness in the number of internal degrees of freedom is nevertheless still strongly constrained by crystalline symmetries, and the investigation of the development of improper orders can lead to very refined information about the nature of order parameters developing in complex materials. 
We expect that these findings can be harvested for a better characterization of ordered phases of matter and in future studies of improper orders of functional materials.

A.S. is grateful for financial support from the Swiss National Science Foundation (SNSF) through Division II (No. 184739). 
A.R. also acknowledges financial support from the Swiss National Science Foundation (SNSF) through an Ambizione Grant No. 186043.

\newpage 
\bibliography{bibliography}

\end{document}